\documentclass[journal=jpclcd,manuscript=letter]{achemso}

\usepackage[version=3]{mhchem} 
\usepackage{graphicx} 
\usepackage{subcaption} 
\usepackage{hyperref}
\usepackage{color}

\mathchardef\mhyphen="2D

\author{Ramón Alain Miranda-Quintana}
\affiliation{Department of Chemistry, University of Florida, Gainesville, FL 32603, USA}
\alsoaffiliation{Quantum Theory Project, University of Florida, Gainesville, FL 32603, USA}
\email{quintana@chem.ufl.edu}
\author{Rugwed Lokhande}
\affiliation{Department of Chemistry, University of Florida, Gainesville, FL 32603, USA}
\alsoaffiliation{Quantum Theory Project, University of Florida, Gainesville, FL 32603, USA}
\author{Carlos E. V. de Moura}
\affiliation{Department of Chemistry, University of Florida, Gainesville, FL 32603, USA}
\alsoaffiliation{Quantum Theory Project, University of Florida, Gainesville, FL 32603, USA}
\author{Krisztina Zsigmond}
\affiliation{Department of Chemistry, University of Florida, Gainesville, FL 32603, USA}
\alsoaffiliation{Quantum Theory Project, University of Florida, Gainesville, FL 32603, USA}
\author{Paul W. Ayers}
\affiliation{Department of Chemistry and Chemical Biology, McMaster University, Hamilton Ontario L8S 4M1, Canada}
\email{ayers@mcmaster.ca}

\title[A new family of seniority-restricted coupled cluster methods]
  {A new family of seniority-restricted coupled cluster methods}

\abbreviations{}

\begin{document}







\begin{abstract}
We introduce a novel class of coupled cluster (CC) methods that leverage the seniority concept to enhance efficiency and accuracy in electronic structure calculations. While existing approaches, such as the pair coupled cluster doubles (pCCD) method, are limited to seniority-zero ($\Omega = 0$) wavefunctions \cite{sg2014}, we propose a more flexible framework: seniority-restricted coupled cluster (sr-CC). This new methodology selectively constrains the seniority sectors accessible through excitation operators in the cluster expansion, enabling a more systematic exploration of electron correlation effects. By balancing computational cost and accuracy, sr-CC provides a promising pathway for advancing electronic structure theory, particularly in strongly correlated systems.
\end{abstract}

\section{Introduction}

Full configuration interaction (FCI) provides an exact solution to the electronic Schrödinger equation within a given basis set by constructing the wavefunction as a linear combination of all possible Slater determinants (or configuration state functions). These configurations are formed by distributing the electrons among the spin orbitals. 
However, its factorial scaling with respect to the number of electrons restricts its practical application to systems containing fewer than approximately 20 electrons. \cite{kp1984, kp1989}
Among the alternative approaches, truncated configuration interaction (CI) and coupled cluster (CC) methods are widely used because they provide systematically improvable approximations to electron correlation effects.\cite{sa1996, ht2013, br2007, br2009}
Truncated CI methods reduce computational cost by limiting the expansion to a subset of  Slater determinants, typically defined by their degree of excitation relative to a reference.
This selective inclusion captures the most significant configurations but sacrifices size extensivity.
By contrast, CC methods utilize an exponential ansatz for the wavefunction, which, even when truncated (e.g., CCSD), implicitly incorporates contributions from higher-order excitations.
This formulation enables CC methods to achieve a favorable balance between accuracy and computational cost while preserving size extensivity.
Typically, these methods begin with a reference configuration---either a single reference (SR methods) or multiple references (MR methods) and progressively introduce electron correlation by incorporating additional determinants from specific subspaces of the Hilbert space.
This is traditionally achieved through an excitation hierarchy.\cite{br2009, ct2000}
The coupled cluster (CC) family of methods represents a robust approach to accurately describe weakly correlated systems.\cite{br1993, br2012, br1981, br1989}
In fact, for systems of moderate size, obtaining highly accurate results is nearly routine.\cite{br2007}
Methods such as coupled cluster with single and double excitations (CCSD)\cite{br1982} and its extension that includes perturbative triple excitations (e.g., CCSD(T))\cite{hgm1989} are widely used for this purpose.
Although this strategy is highly effective for weak (dynamic) correlation, it often fails in strongly correlated regimes because the single-reference approximation breaks down and higher excitations become essential. \cite{br1992, sp2001, br2002} These scenarios commonly occur in bond dissociation processes \cite{br1999, oj2001}, the characterization of radicals \cite{ka2021} and transition metal complexes \cite{hgm2019}.

Therefore, continued innovation in CC techniques tailored to strongly correlated systems remains critical to advancing the many-electron problem.
Pair-coupled cluster doubles (pCCD), a restricted version of coupled cluster doubles (CCD) that only includes excitations preserving electron pairing, exhibit the ability to capture strong electron correlation \cite{bp2013, sg2014_1, mqr2024,tecmerAssessingAccuracyNew2014a,
tecmerGeminalbasedElectronicStructure2022,boguslawskiEfficientDescriptionStrongly2014b}, a capability not typically observed in standard CCD.
In this paper, we further explore the partitioning of Slater determinants using the concept of seniority, where the seniority number ($\Omega$, $\Omega = 0$ in pCCD) counts the number of unpaired electrons in a given electronic configuration.\cite{bytautasSeniorityOrbitalSymmetry2011c}
This alternative partitioning offers a promising path toward more manageable and efficient quantum chemistry methods.

\section{Theory}
\subsection{Seniority Restricted Coupled Cluster}
In coupled cluster theory, the wave function ($ \Psi_{\mathrm{CC}}$) is expanded as an exponential of cluster operators acting on a reference determinant  $|\Phi_0\rangle$, typically the Hartree-Fock determinant.\cite{br2007}
These cluster operators create excitations of electrons from occupied to virtual orbitals, generating correlated wave functions that capture electron correlation effects beyond the mean-field level,  
\begin{equation}
    |\Psi_{\mathrm{CC}}\rangle = e^{\hat{T}} |\Phi_0\rangle,
\end{equation}
where
$\hat{T}$ is the cluster operator, defined as a sum of excitation operators:
    \begin{equation}
        \hat{T} = \hat{T}_1 + \hat{T}_2 + \hat{T}_3 + \ldots 
    \end{equation}

Each excitation operator can be associated with the creation of an electron-hole pair in the reference determinant and is defined as follows:
    \begin{align}
        \hat{T}_1 &= \sum_{ai} t_i^a \hat{a}_a^\dagger \hat{a}_i \\
        \hat{T}_2 &= \frac{1}{4} \sum_{abij} t_{ij}^{ab} \hat{a}_a^\dagger \hat{a}_b^\dagger \hat{a}_j \hat{a}_i  \\
        &\vdots
    \end{align}
In these expressions: $\hat{a}^\dagger$ and $\hat{a}$ denote the second quantization creation and annihilation operators, respectively; occupied orbitals are labeled by indices $i$ and $j$, while the virtual (unoccupied) orbitals are labeled by indices $a$ and $b$;
$t_i^a$, $t_{ij}^{ab}, \ldots$ represent the cluster amplitudes.

The pCCD wavefunction is a $\Omega=0$ coupled cluster wavefunction.\cite{sg2014_1}
In the pCCD approximation, the wavefunction is restricted to include only pairwise excitations.
This means that the excitations occur in pairs from occupied orbitals to virtual orbitals.
In practice, this simplification leads to the following form for the pCCD wavefunction:

\begin{equation}
|\Psi_{\mathrm{pCCD}}\rangle = e^{\hat{T}_2^{\mathrm{pairs}}} |\Phi_0\rangle    
\end{equation}
where \( \hat{T}_2^{\mathrm{pairs}} \) is a constrained version of the cluster operator which only allows paired excitations.
Specifically, instead of summing over all combinations of occupied and virtual orbitals as in CCD, we only consider excitations where electrons are promoted in pairs from a single occupied orbital  ($i\Bar{i}$) to single virtual orbital ($a\Bar{a}$):

\begin{equation}
\hat{T}_2^{\Omega=0} = \hat{T}_2^{\mathrm{pairs}} =\dfrac{1}{4} \sum_{i \Bar{i} a \Bar{a}} t_{i \Bar{i}}^{a \Bar{a} } a_{\Bar{a}}^\dagger a_{{a}}^\dagger a_{\Bar{i}} a_{i}
\end{equation}
Here, in the pair amplitudes, \( t_{i \Bar{i}}^{a \Bar{a} } \), \( i\Bar{i}  \) and  \( a\Bar{a} \) stand for paired excitations (i.e., two electrons from the same spatial orbital).
This restriction reduces the number of excitations and makes the pCCD method computationally more tractable for strongly correlated systems.
The pCCD wavefunction can be improved, after calculating the pair amplitudes, by incorporating single and/or triple excitations.
Including single excitations allows the wavefunction to account for orbital relaxation effects, while triple excitations capture additional dynamic correlation beyond pairwise interactions.
This refinement enhances the accuracy of electronic structure calculations, leading to better energy predictions and a more complete description of electron correlation effects.
These constructions are known as frozen pair CCD (fpCCD) schemes.\cite{sg2014,pandeyFrozenPairTypePCCDBasedMethods2025,
boguslawskiLinearizedCoupledCluster2015,
zobokiLinearizedCoupledCluster2013}

Following this, we introduce a novel class of coupled cluster methods, termed seniority-restricted coupled cluster (sr-CC), which limits accessible seniority sectors by constraining the excitation operators in the cluster operator.
Within this framework, we propose three approaches to construct these wavefunctions.
The simplest variant is the sr-CCSD(0) wavefunction, which includes all single excitations but restricts double excitations to those that preserve a seniority level of zero, ensuring the number of unpaired electrons remains unchanged.
\begin{equation}
|\Psi_{\mathrm{sr\mhyphen CCSD(0)}}\rangle = e^{\hat{T}_1 + \hat{T}_2^{{\Omega=0}}} |\Phi_0\rangle    
\end{equation}

The second approach, sr-CCSDTQ(0), imposes a seniority-zero restriction on the quadruple excitation operator $\hat{T}_4$, while the single ($\hat{T}_1$), double ($\hat{T}_2$), and triple ($\hat{T}_3$) excitation operators remain unrestricted, following the conventional coupled cluster formulation.
\begin{equation}
|\Psi_{\mathrm{sr\mhyphen CCSDTQ(0)}}\rangle = e^{\hat{T}_1 + \hat{T}_2 +\hat{T}_3 +\hat{T}_4^{\Omega=0}} |\Phi_0\rangle    
\end{equation}

\begin{equation}
\hat{T}_4^{\Omega=0} = \dfrac{1}{(4!)^2}\sum_{\substack{i \Bar{i}  {j}\Bar{j} a \Bar{a}b \Bar{b}}}^n t_{i \Bar{i}  {j}\Bar{j}}^{a \Bar{a}b \Bar{b}} a_{\Bar{b}}^\dagger a_b^\dagger a_{\Bar{a}}^\dagger a_a^\dagger a_{\Bar{j}} a_j a_{\Bar{i}} a_i
\end{equation}

Finally, in the sr-CCSDT(2)Q(0) scheme, triple excitations are restricted to seniority two, whereas quadruple excitations are confined to seniority zero.
\begin{equation}
|\Psi_{\mathrm{sr\mhyphen CCSDT(2)Q(0)}}\rangle = e^{\hat{T}_1 + \hat{T}_2 +\hat{T}_3^{\Omega=2} +\hat{T}_4^{{\Omega=0}}} |\Phi_0\rangle    
\end{equation}
where, 
\begin{equation}
\hat{T}_3^{\Omega=2} =  \dfrac{1}{(3!)^2}\sum_{\substack{i \Bar{i}  {j}\\ a \Bar{a}b }}^n t_{i \Bar{i}{j}}^{a \Bar{a}b}  a_b^\dagger a_{\Bar{a}}^\dagger a_a^\dagger  a_j a_{\Bar{i}} a_i
\end{equation}

\begin{figure}[h]
    \centering
    \includegraphics[scale=0.725]{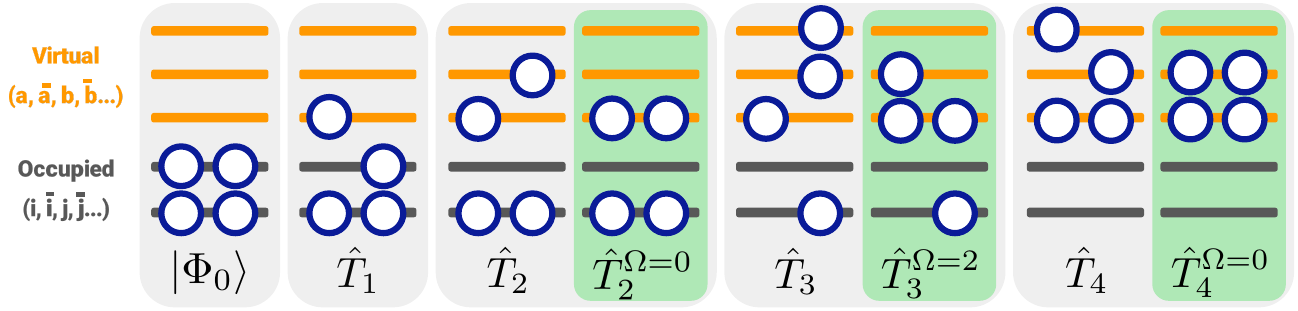}
    \caption{Representation of coupled cluster operators ($\hat{T}$) and paired excitation operators ($\hat{T}^{\Omega}$). Gray and orange bars represent occupied and virtual spatial molecular orbitals, respectively, while white spheres denote electrons.}
    \label{fig:operators}
\end{figure}

In \autoref{fig:operators}, we illustrate the paired excitation operators $\hat{T}^{\Omega}$ used in the sr-CC methods proposed in this paper. These paired excitations can be viewed as a subset of the conventional coupled cluster excitation operators.

All sr-CC methods presented above were implemented using the \textsc{FANCI} framework. 
In this formalism, the wavefunction is expressed as
\begin{equation}
    \big|\Psi(\vec{P})\big) = \sum_{m \in S} f_m(\vec{P}) \, |m\rangle ,
\end{equation}
where $m$ denotes Slater determinants in the chosen space $S$, and $f_m(\vec{P})$ are 
parametric functions corresponding to their overlaps with the ansatz. This \emph{Flexible 
Ansatz for N-body Configuration Interaction} (FANCI) unifies conventional CI and CC 
approaches with tensor-network and geminal-based methods. The parameters are generally 
determined by solving the projected Schrödinger equation,
\begin{equation}
    \sum_{m \in S} f_m(\vec{P}) \, \langle n | \hat{H} | m \rangle 
    = E f_n(\vec{P}), \quad |n\rangle \in P ,
\end{equation}
and the sr-CC methods correspond to specific FANCI parametrizations.\cite{ap2021}

From a computational perspective, the seniority restrictions also reduce the formal 
scaling of the methods. In particular, sr-CCSD(0) scales as 
$\mathcal{O}(N^{4})$, in contrast to the $\mathcal{O}(N^{6})$ cost of 
conventional CCSD. Similarly, both sr-CCSDTQ(0) and sr-CCSDT(2)Q(0) 
exhibit a computational scaling of $\mathcal{O}(N^{6})$, which represents a 
significant reduction relative to the $\mathcal{O}(N^{10})$ scaling of 
full CCSDTQ. These improvements arise because restricting the excitation 
operators to specific seniority sectors eliminates many high-rank tensor 
contractions that dominate the computational cost in conventional coupled 
cluster theory.
\section{Computational Details}

\begin{figure}[h] 
    \centering
    \includegraphics[scale=0.25]{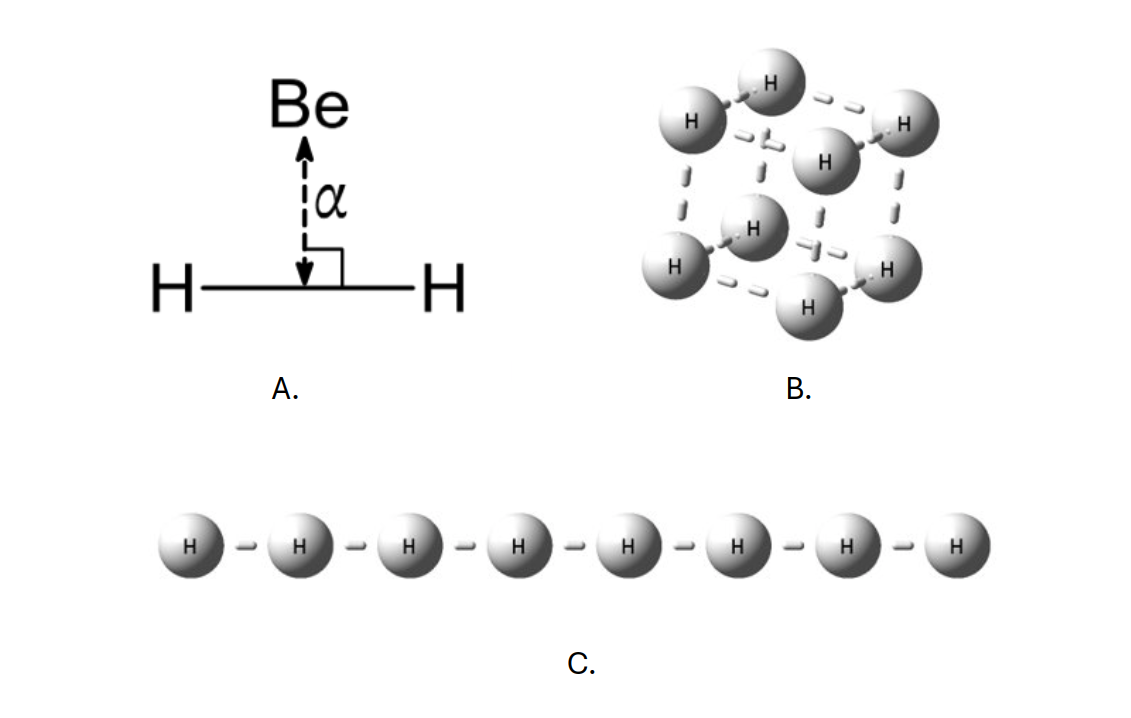} 
    \caption{Representation of model systems: (a) \ce{BeH2}, (b) cubic \ce{H8} and (c) linear \ce{H8}.}
    \label{fig:image1}
\end{figure}

To benchmark the performance of the newly developed sr-CC methods, we evaluated them on three model systems: the  $C_{2v}$  insertion of \ce{Be} into \ce{H2} and the eight-electron hydrogen system \ce{H8} (linear and cubic), as illustrated in \autoref{fig:image1}.

The \ce{BeH2} molecule exhibits a combination of weak and strong electron correlation effects due to the near-degeneracy of its 2s and 2p orbitals.
To investigate these effects, we examined ten molecular geometries based on a previous study by Purvis et al.\cite{pg1983}
Each geometry was generated by varying two key structural parameters: the \ce{H}–\ce{H} bond length and the perpendicular distance between the \ce{Be} atom and the \ce{H2} unit.
For consistency, the \ce{Be} atom was placed at the origin of the coordinate system, and the Cartesian coordinates of the hydrogen atoms for each geometry are listed in \autoref{tab:coordinates}.
The resulting energy profiles illustrate the dependence of the total energy on the perpendicular separation between \ce{Be} and \ce{H2}.

The eight-electron model systems include two configurations: a linear chain of eight equidistant hydrogen atoms and a cubic arrangement.
For the linear \ce{H8} system, a series of geometries was generated by varying the internuclear distance between adjacent hydrogen atoms from 0.5 to 4.0~\AA. In the case of the cubic \ce{H8} system, the edge length of the cube was varied from 0.5 to 4.0 Bohr.
The geometries for this model system were obtained from Ref.\cite{grc2016}

\begin{table}[h!]
\centering
\begin{tabular}{|c|c|c|c|}
\hline
\textbf{Point} & \textbf{X}   & \textbf{Y}          & \textbf{Z}   \\ \hline
A              & 0.0          & $\pm$2.54           & 0.0          \\ \hline
B              & 0.0          & $\pm$2.08           & 1.0          \\ \hline
C              & 0.0          & $\pm$1.62           & 2.0          \\ \hline
D              & 0.0          & $\pm$1.39           & 2.5          \\ \hline
E              & 0.0          & $\pm$1.275          & 2.75         \\ \hline
F              & 0.0          & $\pm$1.16           & 3.0          \\ \hline
G              & 0.0          & $\pm$0.93           & 3.5          \\ \hline
H              & 0.0          & $\pm$0.70           & 4.0          \\ \hline
I              & 0.0          & $\pm$0.70           & 6.0          \\ \hline
J              & 0.0          & $\pm$0.70           & 20.0         \\ \hline
\end{tabular}
\caption{Cartesian coordinates of \ce{BeH2} geometry for different points in Bohr units.}
\label{tab:coordinates}
\end{table}

All methods presented in this work were implemented within a development branch of the Fanpy package\cite{ap2023}, utilizing the FANCI framework to accelerate the implementation and testing of new ab initio approaches\cite{ap2021, ap2024}.
Fanpy is a sandbox open-source Python 3 package specifically developed as a framework for the development and implementation of novel wavefunction ansatzes based on FANCI.\cite{fanpy_github}
Calculations were performed using the STO-6G basis set, with Molecular Orbitals (MOs) derived from Restricted Hartree–Fock (RHF) calculations. The wavefunction parameters were optimized by minimizing a variational energy objective, as implemented in FanPy, using the BFGS quasi-Newton algorithm from the SciPy library.  
The one- and two-body integrals for the Hamiltonian were processed using HORTON.\cite{ap2024_1,kimGBasisPythonLibrary2024}
Reference energy values for the \ce{H8} clusters and \ce{BeH2} model systems were obtained using the Full Configuration Interaction (FCI) method, as implemented in Fanpy. In addition, density functional theory (DFT) calculations with the B3LYP functional, as well as CCSD and CCSD(T), were carried out using the Gaussian 16 program package.\cite{g16}

\section{Results and Discussion}

\begin{figure}[h] 
    \centering
    \begin{subfigure}[b]{0.45\textwidth} 
        \centering
        \includegraphics[scale=0.335]{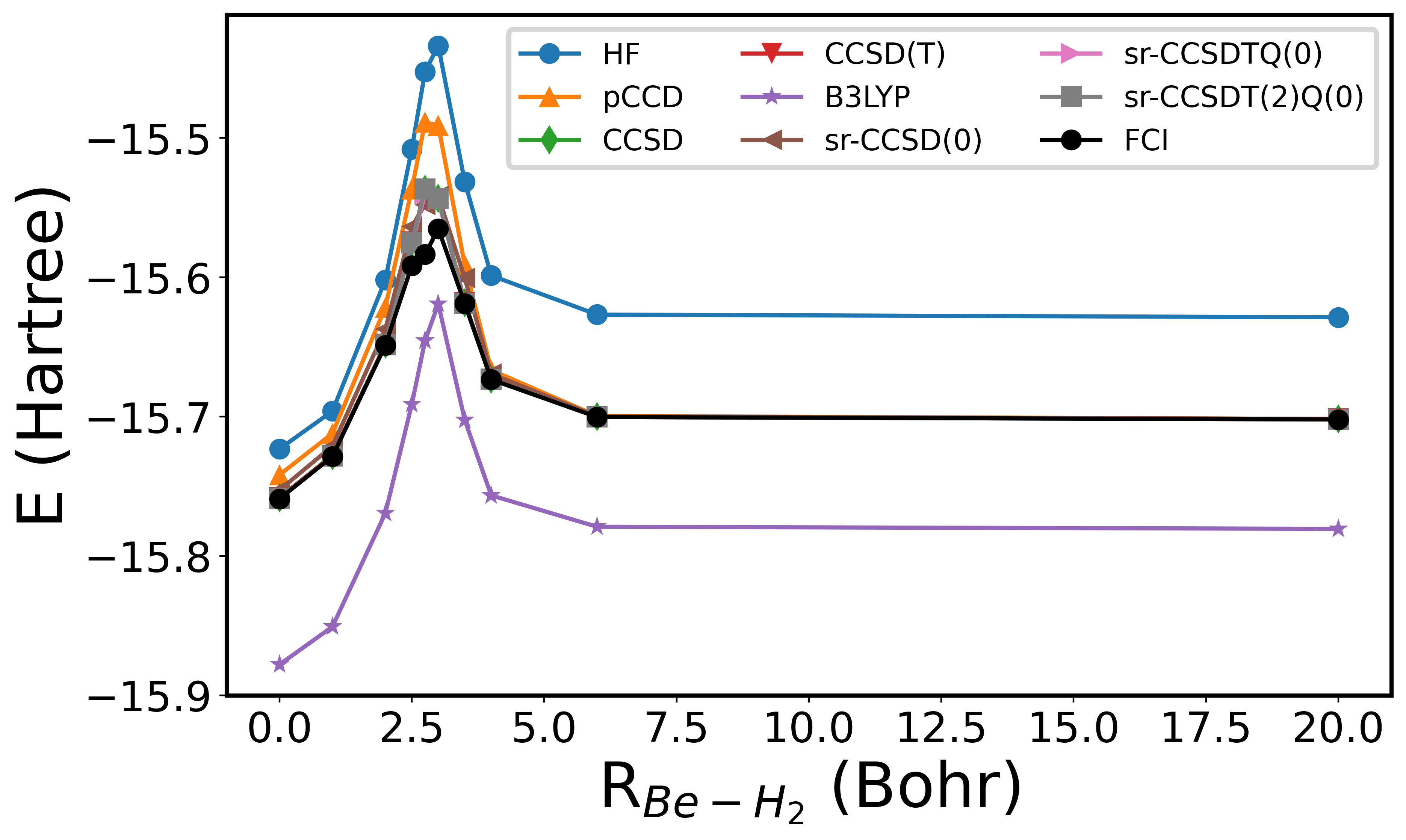} 
        \caption{} 
        \label{fig:image4}
    \end{subfigure}
    \hfill
    \begin{subfigure}[b]{0.45\textwidth} 
        \centering
        \includegraphics[scale=0.335]{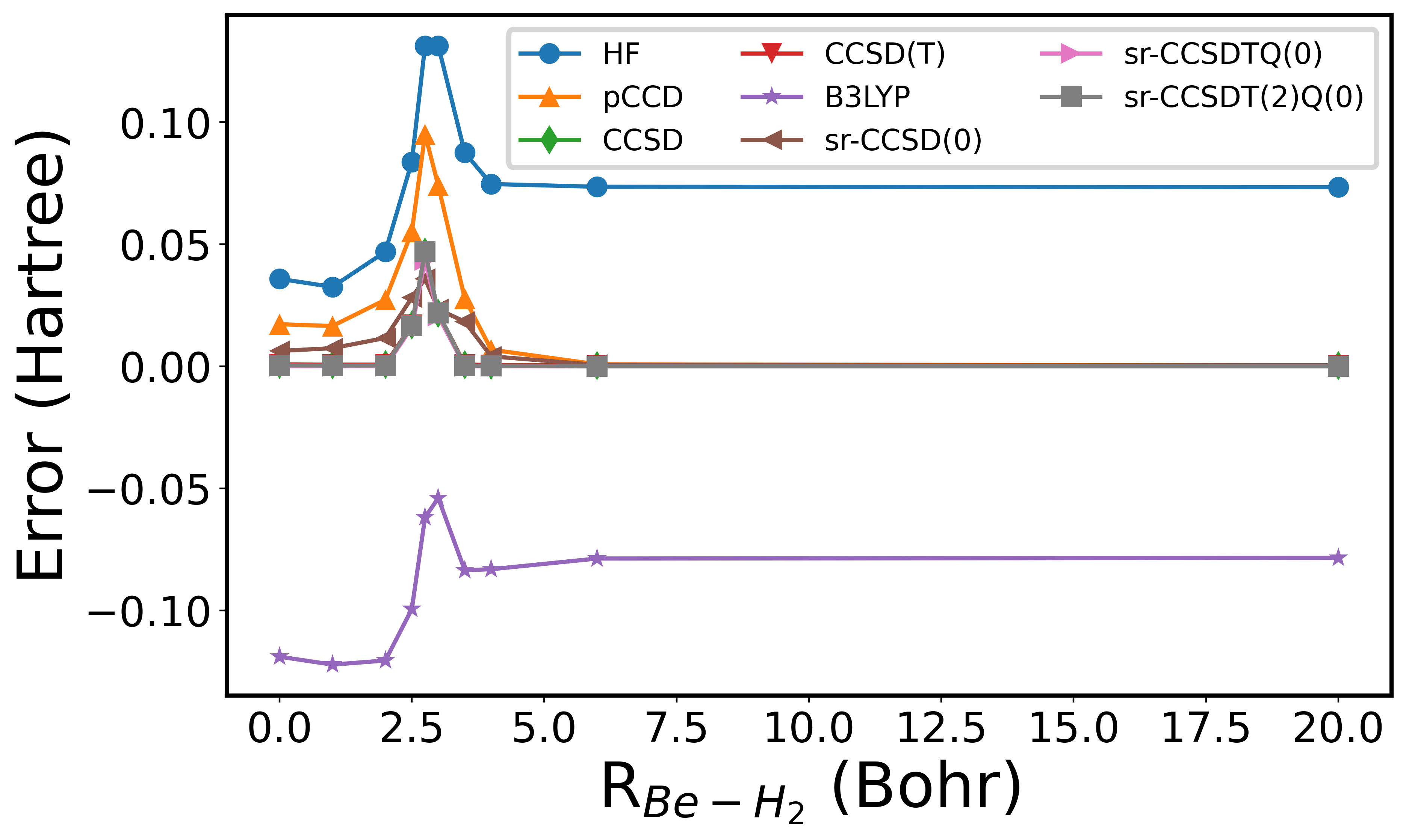} 
        \caption{} 
        \label{fig:image5}
    \end{subfigure}
    
    \caption{Potential energy curves of $C_{2v}$ insertion of \ce{Be} into \ce{H2}. (a) Total energies and (b) energy errors relative to FCI, computed using the sr-CC methods. All results obtained using the STO-6G basis set.}
    \label{fig:results-BeH2}
\end{figure}

We present the results of our investigation for the ground state energies of the \ce{BeH2} and linear and cubic \ce{H8} systems, calculated using the newly developed sr-CC wavefunctions, in comparison to pCCD wavefunction.
In \autoref{fig:results-BeH2}, we display the potential energy curves for \ce{BeH2} along the $C_{2v}$ insertion surface, parameterized by $\alpha$. The results show a clear improvement in the energy predictions when using sr-CC methods compared to the pCCD method.
The sr-CC methods exhibit exceptional agreement with the reference FCI data. While pCCD is limited to pair-preserving doubles and typically requires explicit orbital optimization to achieve comparable accuracy, the sr-CC ansatz incorporates singles directly in the cluster expansion. This additional flexibility yields significantly better agreement with the FCI reference without the need for orbital rotations.  For comparison, we also examined conventional DFT (B3LYP) and standard excitation based CC methods including CCSD and CCSD(T). B3LYP substantially overestimates the total energy across the insertion curve, reflecting the inability of approximate functionals to treat near-degenerate orbitals of Be and the multi-configurational character of the system. In contrast, CCSD and CCSD(T) perform well, essentially reproducing the FCI reference and yielding results comparable to sr-CCSD(0). Thus, for this moderately correlated case, excitation-based CC methods remain reliable, but DFT fails to capture the essential physics. 

This finding corroborates the expectation that methods incorporating higher-order excitations and more parameters, such as sr-CCSDTQ(0) and sr-CCSDT(2)Q(0), provide more precise results than their lower-order counterparts. However, at mid-range geometries ($R_{\ce{BeH2}}$ = 2.5, 2.75, 3.0 a.u.), the energy error reaches an order of 0.04 Hartrees for  sr-CCSDTQ(0) and sr-CCSDT(2)Q(0) and 0.03 Hartrees for sr-CCSD(0) method, at $R_{\ce{BeH2}}$ = 2.75 a.u. In this regime, significant orbital mixing occurs between the Be 2s/2p orbitals and the hydrogen 1s orbitals, leading to partial bond formation and delocalization of electron density across all three atoms. This situation creates a multiconfigurational electronic structure with strong static and dynamic correlation effects that cannot be easily captured by lower-order or single-reference methods. As the internuclear separation increases, the errors diminish significantly, with the energy error approaching a value as small as 0.2 mHartrees for sr-CCSD(0) and negligible errors for sr-CCSDT(2)Q(0) and sr-CCSDTQ(0).

\begin{figure}[h] 
    \centering
    \begin{subfigure}[b]{0.45\textwidth} 
        \centering
        \includegraphics[scale=0.335]{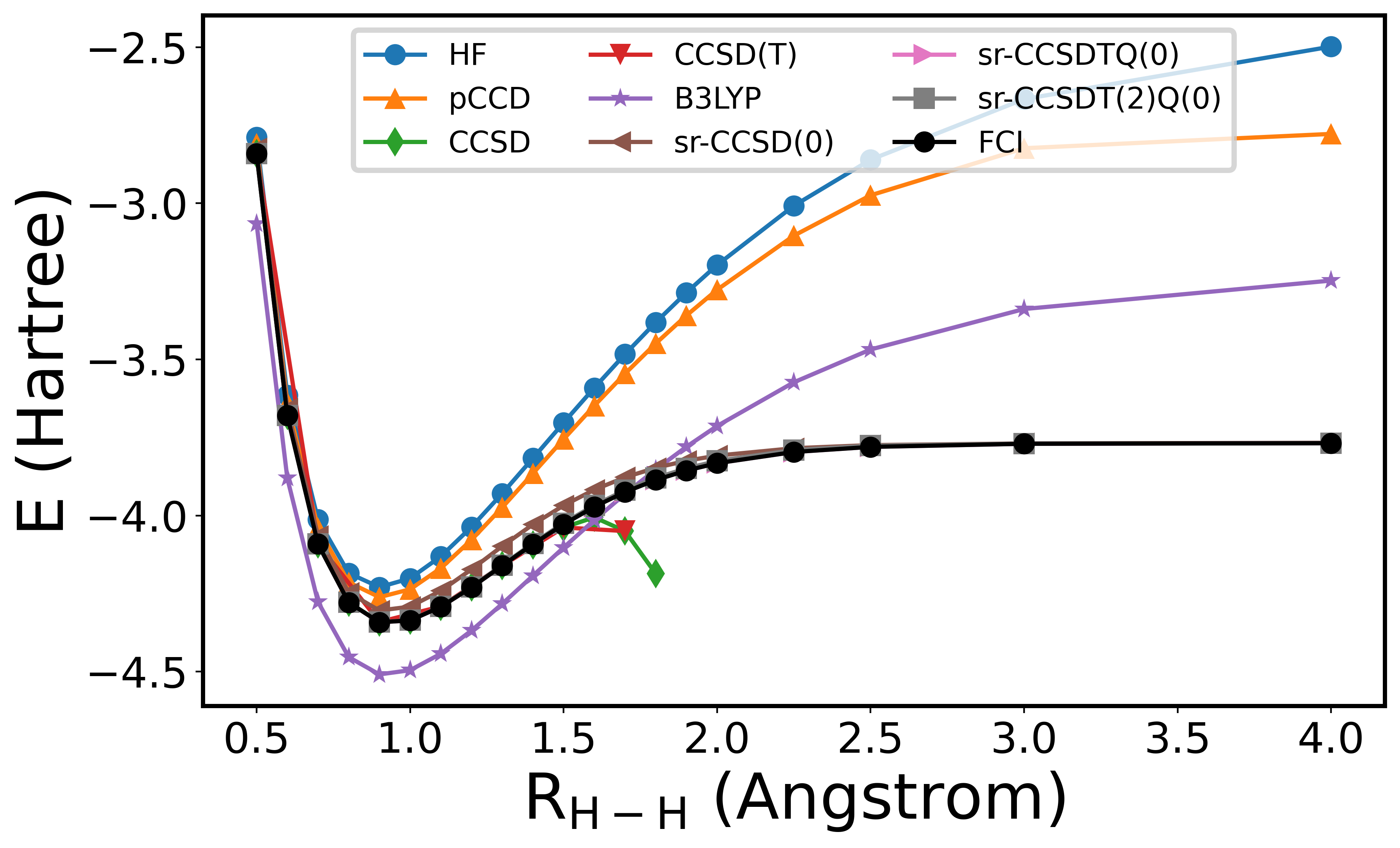} 
        \caption{} 
        \label{fig:image6}
    \end{subfigure}
    \hfill
    \begin{subfigure}[b]{0.45\textwidth} 
        \centering
        \includegraphics[scale=0.335]{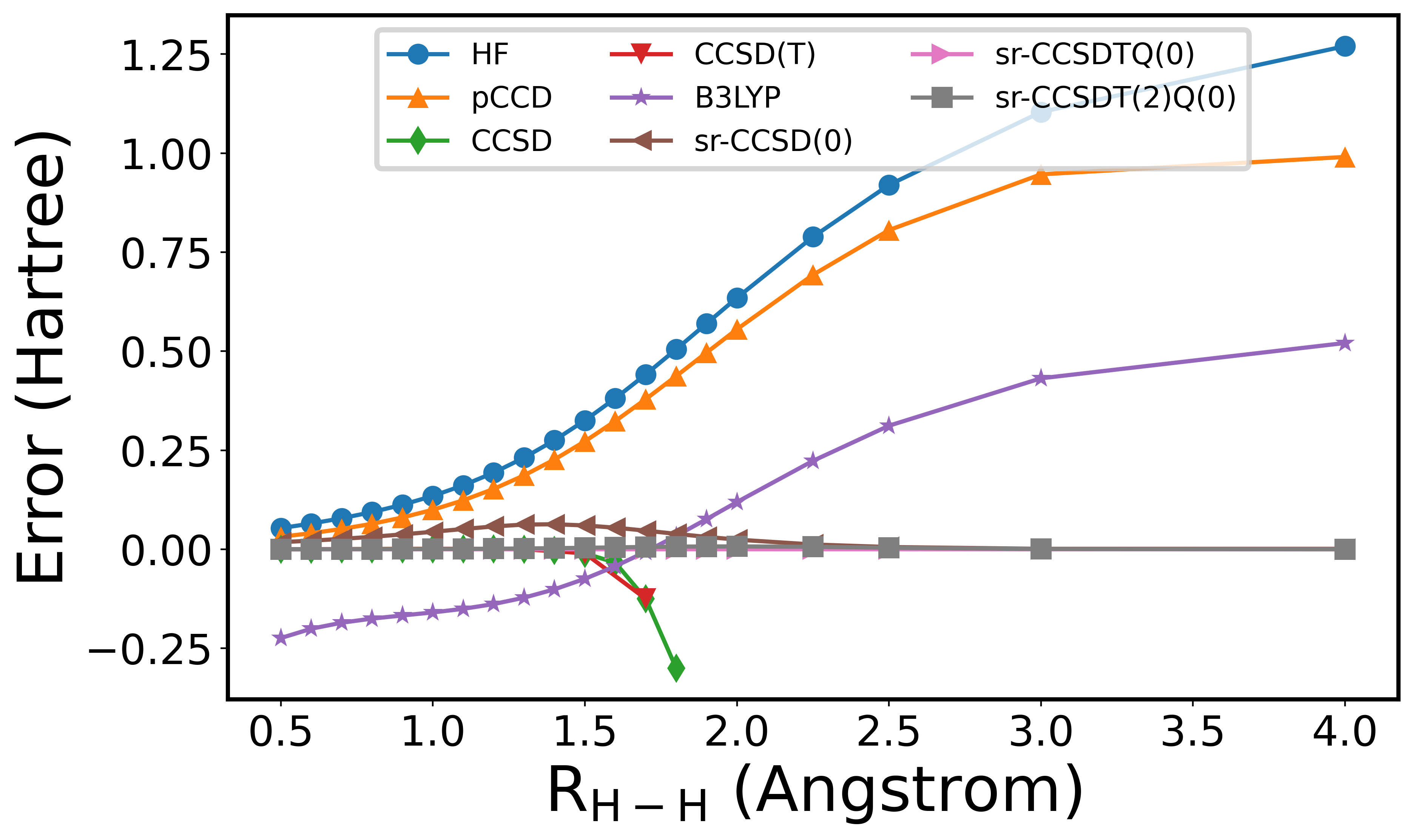} 
        \caption{} 
        \label{fig:image7}
    \end{subfigure}
    
    \caption{Potential energy curves for linear \ce{H8}. (a) Total energies and (b) energy errors relative to FCI, computed using the sr-CC methods. All results obtained using the STO-6G basis set.}
    \label{fig:results-H8}
\end{figure}

We further investigate the performance of the sr-CC methods on a larger system, linear \ce{H8}, which involves a greater number of electrons and a larger computational system size.
\autoref{fig:results-H8} presents the potential energy curves for the symmetric dissociation of linear \ce{H8}, comparing the sr-CC, HF (Hartree-Fock), and FCI results. For the linear \ce{H8} chain, conventional methods perform much worse. B3LYP strongly overestimates the total energy along the dissociation curve because of delocalization and self-interaction errors. CCSD also breaks down as the H–H distance increases, and CCSD(T) performs even worse since the perturbative triples correction diverges in regions with strong multireference character.
In the case of sr-CCSD(0), we observe a significant lack of correlation for H-H separations between 0.8 \AA  \  and 2.0 \AA \  with errors up to 0.06 Hartrees.
However, for longer bond distances and in the dissociation region, sr-CCSD(0) aligns well with the FCI results, demonstrating its capability to capture the essential physics of bond dissociation, a domain where traditional coupled-cluster methods often fail.
In contrast, sr-CCSDT(2)Q(0) and sr-CCSDTQ(0) methods deliver nearly perfect agreement with FCI over the entire dissociation range, demonstrating their superior accuracy.
The error analysis further corroborates this finding, with sr-CCSDT(2)Q(0) showing errors on the order of $10^{-3}$, while sr-CCSDTQ(0) achieves remarkable precision, with errors on the order of $10^{-5}$ mHartrees.

\begin{figure}[h] 
    \centering
    \begin{subfigure}[b]{0.45\textwidth} 
        \centering
        \includegraphics[scale=0.335]{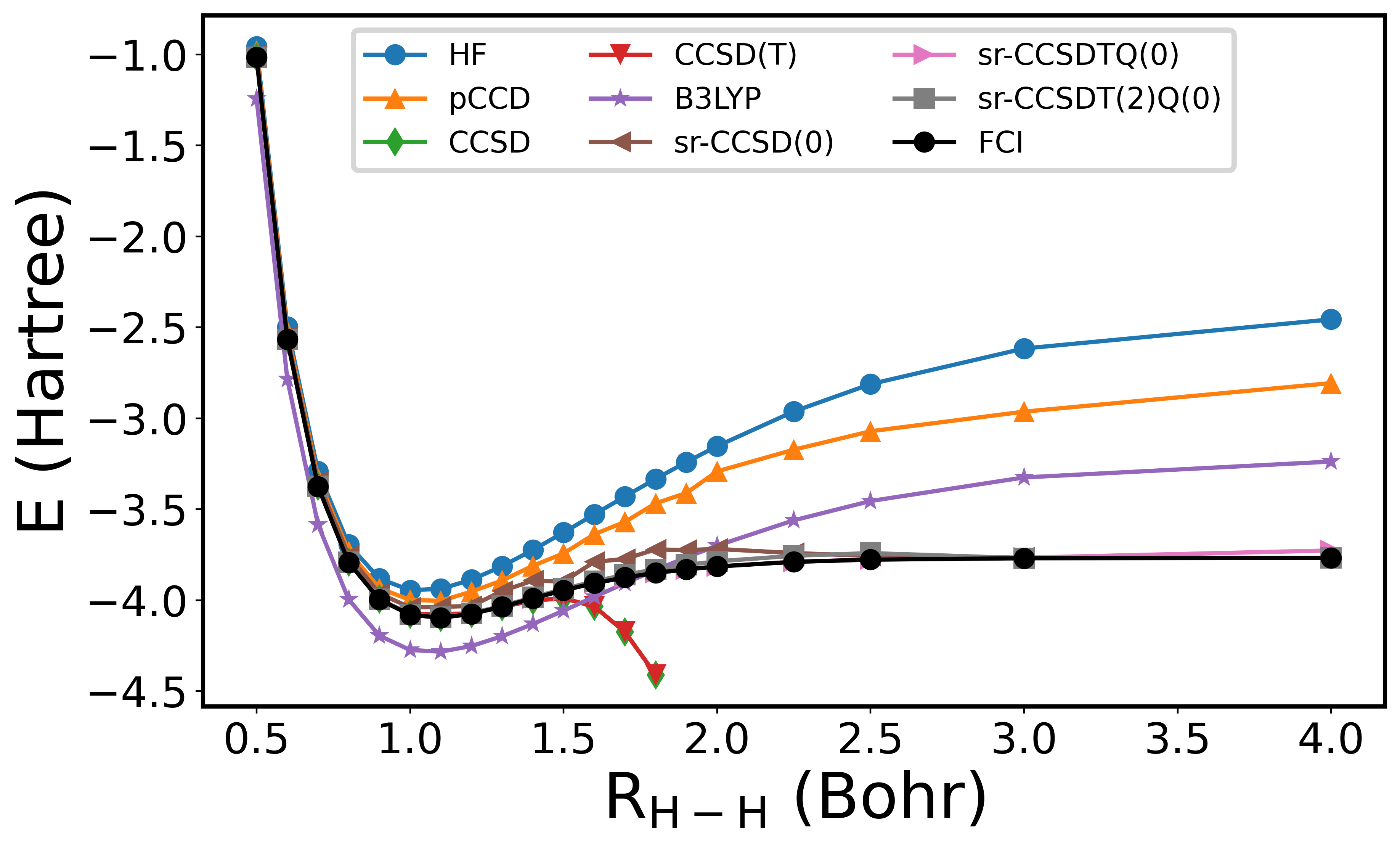} 
        \caption{} 
        \label{fig:image8}
    \end{subfigure}
    \hfill
    \begin{subfigure}[b]{0.45\textwidth} 
        \centering
        \includegraphics[scale=0.335]{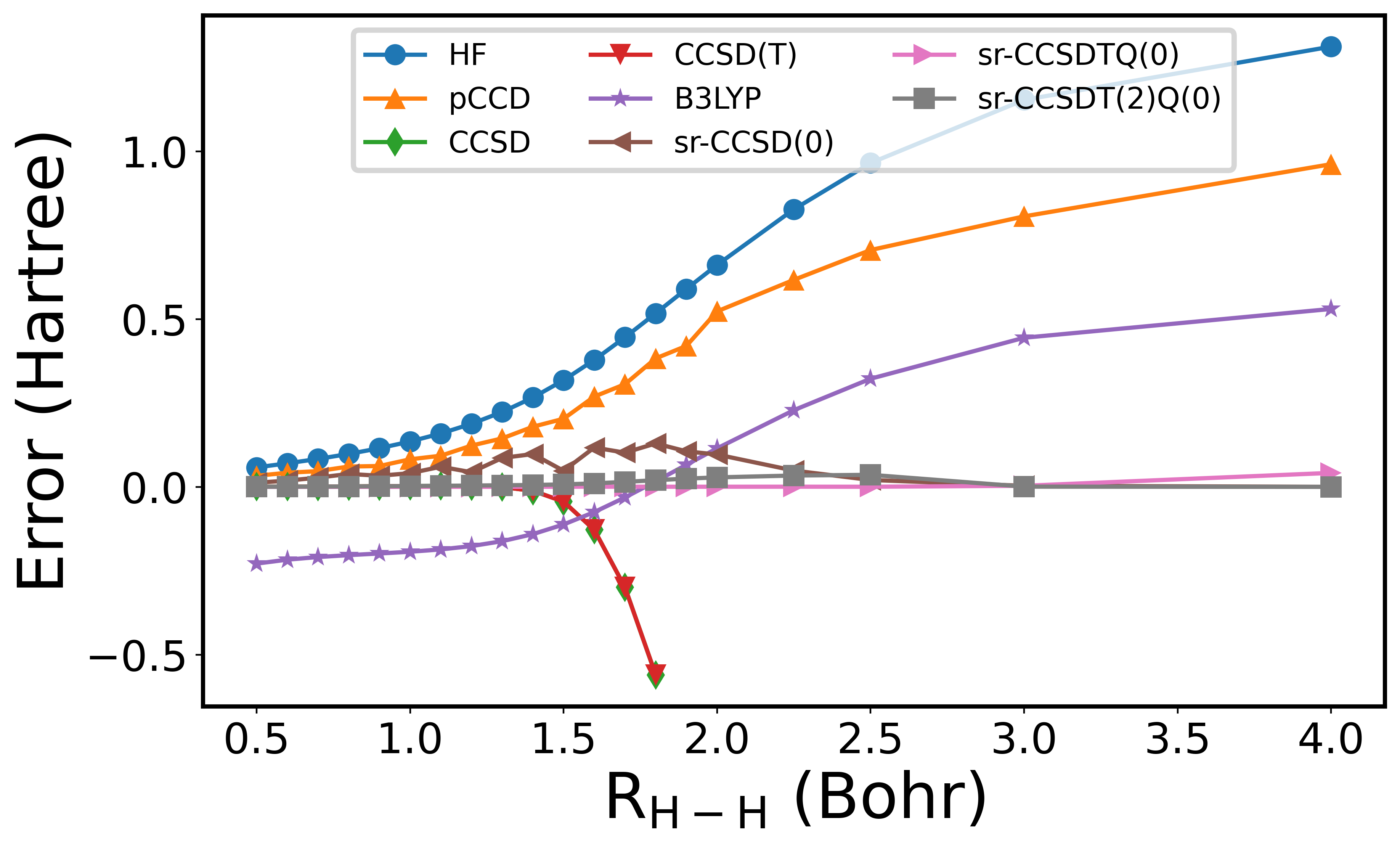} 
        \caption{} 
        \label{fig:image9}
    \end{subfigure}

    \caption{Potential energy curves for cubic \ce{H8}. (a) Total energies and (b) energy errors relative to FCI, computed using the sr-CC methods. All results obtained using the STO-6G basis set.}
    \label{fig:results-cubic-H8}
\end{figure}

In \autoref{fig:results-cubic-H8}, we examine the cubic \ce{H8} system. Both CCSD and CCSD(T) fail, producing unphysical results, while B3LYP substantially overestimates the energies. The sr-CCSD(0) method shows convergence difficulties, leading to numerical instabilities that appear as kinks in the potential energy curve, suggesting that the sr-CCSD(0) method is still useful for obtaining rough energy estimates.
However, the sr-CCSDTQ(0) and sr-CCSDT(2)Q(0) methods show smooth, stable performance, closely following the reference FCI data with maximum errors on the order of $10^{-4}$ and $10^{-2}$ Hartrees, respectively.
Furthermore, across all systems investigated, the new sr-CC methods consistently outperform pCCD in the presence of single excitations.
This highlights their potential to circumvent the need for orbital rotations, an advantage recently emphasized in studies by Gaikwad, thereby providing a more direct and efficient route to accurate wavefunction representations.\cite{mqr2024}
The sr-CC wavefunctions, which incorporate single-reference correlation treatment, exhibit superior performance in capturing both electron correlation and bond dissociation dynamics. 

\newpage
\subsection{Conclusions}

In this work, we introduce the seniority-restricted coupled cluster (sr-CC) approach, a novel method that constrains the accessible seniority sectors in coupled cluster theory.
This framework offers an efficient way to handle electron correlation, particularly in systems with strong correlation effects, which are difficult to capture with conventional CC methods.

We propose three variants of the sr-CC method, each imposing different seniority restrictions on the excitation operators.
The first variant, sr-CCSD(0), includes all single excitations while restricting double excitations to those that preserve a seniority level of zero, ensuring that the number of unpaired electrons remains unchanged.
The second variant, sr-CCSDTQ(0), allows unrestricted single, double, and triple excitations but constrains the quadruple excitation operator to seniority zero. 
The sr-CCSDT(2)Q(0) variant imposes a seniority-two restriction on triple excitations, while quadruple excitations remain restricted to seniority zero.

Our computational results for \ce{BeH2} system reveal that the sr-CC method provides a significant improvement over the pCCD method, especially in capturing both weak and strong electron correlation effects.
In particular, sr-CCSDTQ(0) and sr-CCSDT(2)Q(0) closely follow the reference FCI data, significantly outperforming sr-CCSD(0), which shows a lower accuracy at both very short and long bond lengths.

For the \ce{H8} systems, both linear and cubic configurations, the sr-CC methods show consistent and accurate energy predictions when compared to the FCI results, validating their robustness and reliability.
The methods provide a balanced treatment of both dynamic and strong electron correlation, making them suitable for a wider range of molecular systems.

Overall, the sr-CC methods represent a promising new approach in quantum chemistry, combining efficiency with the ability to tackle strongly correlated systems.
Future developments could involve extending these methods to even more complex systems, refining the incorporation of higher-order excitations, further optimizing their computational performance for larger-scale systems, and exploring whether the accuracy of these approaches could, like sr-CCD(0), be improved by including orbital optimization.\cite{boguslawskiProjectedSenioritytwoOrbital2014b,
limacherInfluenceOrbitalRotation2014} 

\begin{acknowledgement}
With this manuscript we want to honor the memory of the late John F. Stanton. A superb friend, colleague, and mentor, John deeply influenced our work through his boundless optimism and encyclopedic and out-of-the-box chemical knowledge. We acknowledge support from the National Science Foundation CAREER award CHE-2439867 (RAMQ), NSERC, the Canada Research Chairs, and the Digital Research Alliance of Canada (PWA).
\end{acknowledgement}


\bibliography{achemso-demo}

\end{document}